\documentclass[journal=jpcafh,
               manuscript=article,
               ]{achemso}

\usepackage{hyperref}

\usepackage{amsmath}
\usepackage{graphicx}
\usepackage{tabularx}
\usepackage[usenames,dvipsnames]{xcolor}
\usepackage[version=3]{mhchem}
\usepackage{makecell}

\newcounter{pabh}
\newcounter{pabhq}
\newcounter{update}
\newcounter{ephraim}
\newcounter{anastasia}

\definecolor{ao}{rgb}{0.0, 0.5, 0.0}

\newcommand{\qed}{QED$^{\text{VP+SE}}$ }

\title{Hyperfine structure constants on the relativistic coupled cluster level with associated uncertainties}

\author{Pi A. B. Haase}
\affiliation{Van Swinderen Institute, University of Groningen, 9747 Groningen, The Netherlands}
\email{p.a.b.haase@rug.nl}

\author{Ephraim Eliav}
\affiliation{School of Chemistry, Tel Aviv University, 69978 Tel Aviv, Israel}

\author{Miroslav Ilia{\v{s}}} 
\affiliation{Department of Chemistry, Faculty of Natural Sciences, Matej Bel University,Tajovsk{\`e}ho 40 , SK-97400 Banska Bystrica, Slovakia}

\author{Anastasia Borschevsky}
\affiliation{Van Swinderen Institute, University of Groningen, 9747 Groningen, The Netherlands}

\begin{document}

%
%
%
%


\maketitle

\section*{Abstract}

Accurate predictions of hyperfine structure (HFS) constants are important in many areas of chemistry and physics, from the determination of nuclear electric and magnetic moments to benchmarking of new theoretical methods. We present a detailed investigation of the performance of the relativistic coupled cluster method for calculating HFS constants withing the finite-field scheme. The two selected test systems are $^{133}$Cs and $^{137}$BaF. Special attention has been paid to construct a theoretical uncertainty estimate based on investigations on basis set, electron correlation and relativistic effects. The largest contribution to the uncertainty estimate comes from higher order correlation contributions. Our conservative uncertainty estimate for the calculated HFS constants is $\sim$ 5.5\%, while the actual deviation of our results from experimental values was $<1$\% in all cases. 

\section{Introduction}

The hyperfine structure (HFS) constants parametrize the interaction between the electronic and the nuclear electromagnetic moments. 
The HFS consequently provides important information about the nuclear as well as the electronic structure of atoms and molecules and can serve as a fingerprint of, for example, transition metal complexes, probed by electron paramagnetic resonance (EPR) spectroscopy\cite{Abragam1970}, or of atoms, ions, and small molecules in the field of atomic and molecular physics, investigated by optical or microwave spectroscopy. With the ever relentless progress in the field of atomic and molecular precision experiments, there is a growing need for both experimental and theoretical determination of the HFS. 
Accurate calculations of the HFS parameters can serve a \textit{direct} as well as an \textit{indirect} purpose as will be elaborated in the following.

One example of a direct application of accurate theoretical HFS parameters is  nuclear studies, where the calculated electronic properties (magnetic induction and electric field gradient) are used to extract the nuclear magnetic dipole and electric quadrupole moments of the heaviest or unstable atomic nuclei from the measured magnetic-dipole, $A$, and electric-quadrupole, $B$, HFS constants, respectively\cite{Ferrer2017,Raeder2018}. Another example is in the search for even better atomic clocks where the structure of the hyperfine levels must be known to great accuracy in order to make reliable predictions to guide new experiments\cite{Kozlov2018}.

 The calculated values of the HFS constants can be also used as a means to benchmark the employed theoretical method against existing experimental or higher level theoretical data. In order for a theoretical method to yield accurate predictions of the HFS constants, the electron distribution in the vicinity of the atomic nucleus in question must be properly described; comparison to experiment can thus give an indication of the quality of the employed wave function. Such applications can be considered to serve an indirect purpose. 

Using HFS constants as benchmarks is particularly valuable when one is interested in a property that is sensitive to the interaction between electrons and nuclei and that can not be obtained experimentally. One such example is the interpretation of the atomic parity nonconserving (PNC) measurements in Cs atoms, where theoretically determined PNC matrix elements are needed in order to extract the weak charge, i.e. the strength of the neutral weak interaction, from the measured transition amplitudes\cite{Wood1997, Bennett1999}. These matrix elements are sensitive to relativistic effects, which become important when the electrons are close to the atomic nucleus. Therefore, the accuracy of the calculated HFS constants (compared to experiment)  serves as a good indication of the reliability of the predictions for the PNC matrix elements. In order to unambiguously test agreement with the Standard Model prediction of the weak charge, the uncertainty of the theoretical predictions needed to be smaller than 1\%; such accuracy eventually was reached by several groups using calculated HFS constants as benchmark values\cite{Blundell1991, Safronova1999, Kozlov2001, Derevianko2001, Porsev2010, DzuBerFla12}. Such system-specific sensitivity or enhancement factors are generally needed in the search for physics beyond the Standard Model with atoms and molecules\cite{GinFla04, SafBudDeM18, DeMille2015}. 

When accurate predictions of the HFS constants for heavy atoms or for molecules containing heavy elements are needed, special attention must be paid to two aspects: relativistic effects and electron correlation. In addition, it is desirable to use a method that allows reliable uncertainty estimates. In this study we investigate a scheme that meets these three requirements.

In the rest of this paper we will consider the magnetic-dipole HFS constant, which we will refer to as simply the HFS constant. We begin with an overview of the currently popular methods used in the calculations of this property.

 As we are interested in high accuracy treatment of correlation and relativistic effects, we will limit this overview to methods that treat relativity beyond scalar relativistic effects and correlation beyond density functional theory (DFT). For an overview of nonrelativistic as well as DFT based methods we refer to the chapter by H. Bolvin and J. Autschbach\cite{Bolvin2017}. For atoms, methods such as the multi-configurational Dirac-Fock (MCDF)\cite{Jonsson2007}, Dirac-Hartree-Fock augmented by the many body perturbation theory (DHF + MBPT)\cite{Dzuba1984, Das1987}, configuration interaction with MBPT (CI + MBPT)\cite{Dzuba1996,Dzuba1998}, all order correlation potential\cite{Dzuba1987}, coupled cluster singles doubles with partial triples (SDpT)\cite{Safronova1998,Safronova1999} as well as Fock-space coupled cluster (FSCC)\cite{Das2011} were shown to provide reliable results. For molecules, the situation becomes more complicated due to the lack of spherical symmetry and a limited number of implementations exist. These include the multi-reference configuration interaction (MR-CISD) method\cite{Fleig2014a}, the restricted active space CI (RAS-CI) approach \cite{Sasmal2015}, as well as the coupled cluster singles and doubles (CCSD) method\cite{Sasmal2015,Sasmal2017}.

In this work we investigate the performance of the relativistic coupled cluster (CC) method for calculating the HFS constants of atoms and molecules.
Where applicable, this approach provides the highest level of theory, while still being feasible for computations on the heaviest elements. In addition, the systematic construction of the CC method allows for a reliable uncertainty estimation. In this work we combine the CC approach with the well-known finite field scheme (also known as the finite difference method) to extract the HFS constants. This provides us with a straightforward way to calculate molecular properties as numerical derivatives\cite{Cohen1965}. The finite field approach is particularly useful in the framework of the CC theory, since the formulation of expectation values is cumbersome due to the complicated form of the wave function. That said, several implementations exist for calculation of CC expectation values; the recent relativistic examples are the extended CC method (ECC)\cite{Sasmal2015}, the Z-vector CC method\cite{SASMAL2016}, and analytic gradients approach\cite{Shee2016}. An advantage of using the finite field method is that no truncation of the CC expansion is necessary (which is the case for the ECC method for example) and that it allows inclusion of the perturbative triple excitations without additional complications. A drawback of the finite field method is the increased computational cost. Furthermore, one has to pay special attention to the numerical stability.

The combination of the relativistic CC method and the finite field approach has previously been applied to various properties, such as
dipole polarizabilities\cite{Lim2004}, electric field gradients\cite{Arcisauskaite2012,Visscher1998,YakEliKal07}, contact densities for calculating M\"{o}ssbauer isomer shifts\cite{Knecht2011} and $P$- and $P,T$-odd relativistic enhancement factors\cite{Abe2018, Hao2018,Denis2019}. The combination of the CC method and the finite field approach for calculating HFS constants has previously been used in a non-relativistic framework\cite{Sekino1985,Bartlett1979,Carmichael1990}, but, to the best of our knowledge,  the extension to a relativistic framework and application to systems with heavy atoms have not been demonstrated before. Here, we investigate the performance of this method and the effect of various computational parameters (e.g. basis set quality, active space size, treatment of higher order relativistic effects, and others) on the obtained results. Furthermore, we employ a straightforward and reliable scheme for assigning uncertainties of the calculated HFS constants.

Inspired by the examples mentioned above we have chosen to apply our investigations to the HFS constants of the Cs atom and the BaF molecule. Due to the atomic PNC experiments, the HFS constant of Cs has been studied extensively and on high levels of theory, which makes it an ideal system for benchmark calculations. The BaF molecule is currently used in various experiments searching for physics beyond the Standard Model\cite{Vutha2017,Altuntas2018,Aggarwal2018}, where theoretically determined enhancement factors are crucial for the interpretation of the measurements and the calculated HFS constants can provide an important indication of the theoretical uncertainty.

\section{Theory}\label{sec:theory}

The magnetic hyperfine interaction between the electronic spin and the nuclear spin of the $M$th nucleus is parametrized by the 3x3 hyperfine coupling tensor, $\mathbf{A}^M$. It is usually defined through the effective spin Hamiltonian\cite{Abragam1951}:
\begin{equation}\label{eq:spin_H}
H^{M,\text{HFS}}_{\text{spin}}= \vec{I}^M \mathbf{A}^M \vec{\tilde{S}} 
= \sum_{uv} I_u^M A_{uv}^M \tilde{S}_v ,
\end{equation}
where $\vec{\tilde{S}}$ is the effective electronic spin operator and $\vec{I}^M$ is the spin of nucleus $M$. The expectation value of this operator over pure spin-functions, with spin quantization along the $v$-axis, gives the energy due the hyperfine interaction:
\begin{equation}
E^{(v)}_{\text{spin}}(\vec{I}^M) = \sum_u I_u^M A_{uv}^M \langle \tilde{S}_v \rangle.
\end{equation}
This energy will be equal to the true hyperfine interaction energy\cite{Pryce1950,Griffith1960,McWeeny1965}, obtained via a quantum mechanical description, $E^{(v)}_{\text{QM}}(\vec{I}^M)$. In other words, the result for the effective spin Hamiltonian can be mapped onto the results of the quantum mechanical Hamiltonian\cite{Bolvin2017}.
In order to determine an element of the hyperfine coupling tensor, the derivative with respect to the $u$th component of the nuclear spin is taken:
\begin{equation}\label{eq:deriv}
A_{uv}^M = \frac{1}{\langle \tilde{S}_v \rangle} \frac{\text{d} E_{\text{QM}}^{(v)}(\vec{I}^M)}{\text{d} I^M_u}.
\end{equation}

In the following, an appropriate quantum mechanical operator describing the hyperfine interaction will be derived starting from the relativistic Dirac Hamiltonian, with the electron-electron interaction given by the Coulomb operator\cite{SAUE1997}:

\begin{equation}\label{eq:dirac}
\hat{H} = \sum_i \left[
(\boldsymbol{\beta}_i - 1) c^2
+ c \vec{\boldsymbol{\alpha}}_i \cdot \hat{\vec{p}}_i
+ V_{\text{nuc}}(i)
\right]
+ \frac{1}{2} \sum_{i\ne j} \frac{1}{r_{ij}},
\end{equation}
where $\vec{\boldsymbol{\alpha}}$ and $\boldsymbol{\beta}$ are the Dirac matrices:
\begin{equation}
\vec{\boldsymbol{\alpha}} = 
\begin{pmatrix}
0 & \vec{\boldsymbol{\sigma}} \\
\vec{\boldsymbol{\sigma}} & 0 \\
\end{pmatrix}
, ~~
\boldsymbol{\beta} = \begin{pmatrix}
\mathbf{1} & 0 \\
0 & \mathbf{-1}
\end{pmatrix},
\end{equation}
and $\vec{\boldsymbol{\sigma}}$ is the vector consisting of the Pauli spin matrices:
\begin{equation}
\boldsymbol{\sigma}_x = 
\begin{pmatrix}
0 & 1 \\
1 & 0 
\end{pmatrix}
, ~~
\boldsymbol{\sigma}_y = 
\begin{pmatrix}
0 & -i \\
i & 0 
\end{pmatrix}
, ~~
\boldsymbol{\sigma}_z = 
\begin{pmatrix}
1 & 0 \\
0 & -1 
\end{pmatrix}.
\end{equation} 
The nuclear potential in Eq. (\ref{eq:dirac}), $V_{\text{nuc}}(i)$, is approximated by a finite nuclear charge distribution in the shape of a Gaussian function\cite{Visscher1997b}.

To derive the operator for the hyperfine interaction, the magnetic field from the $M$th nucleus is introduced in the Dirac Hamiltonian via the minimal coupling (using the cgs system of atomic units)\cite{Schwartz1955}:
\begin{equation}
\vec{p} \rightarrow \vec{p} + \frac{1}{c} \vec{A}^M(\vec{r}_i),
\end{equation}
where $\vec{A}^M$ is the vector potential; within a point-like description of the magnetization distribution it is given by 
\begin{equation}
\vec{A}^M(\vec{r}_i) = \frac{\vec{\mu}^M \times \vec{r}_{iM}}{r_{iM}^3},
\end{equation}
where $\vec{\mu}^M$ is the magnetic moment of nucleus $M$ given by $\vec{\mu}^M=g_M\mu_{\text{N}}\vec{I}^M$, with $g_M$ the nuclear $g$-factor and $\mu_{\text{N}}$ the nuclear magneton ($\mu_N=(2m_pc)^{-1}$).

Keeping only the term including $\vec{A}^M$ gives the one-electron hyperfine interaction operator:
\begin{equation}
\hat{H}^{M,\text{HFS}} = \sum_i \boldsymbol\alpha_i \cdot \vec{A}^M(\vec{r}_i),
\end{equation}
and inserting the expression for the vector potential yields:
\begin{eqnarray} 
\hat{H}^{M,\text{HFS}} &=& g_M \mu_{\text{N}} \vec{I}^M \cdot \sum_i \frac{( \vec{r}_{iM} \times \vec{\boldsymbol{\alpha_i}} )}{r_{iM}^3} \\
 &=& \sum_{u} g_M \mu_{\text{N}} I^M_u \sum_i \frac{( \vec{r}_{iM} \times \vec{\boldsymbol{\alpha_i}} )_u}{r_{iM}^3} \\
 &=& \sum_{u} I^M_u \hat{H}^{M,\text{HFS}}_u  \label{eq:qm_operator}. \label{eq:hfs_op}
\end{eqnarray} 

In the case of variational wave functions (such as Hartree-Fock, DFT, CI, etc.) the derivative in Eq. (\ref{eq:deriv}) can be translated into an expectation value using the Hellmann-Feynman theorem. In this work we employ the finite field method\cite{Cohen1965}, where the derivative is evaluated numerically. In the finite field method the perturbation operator is added to the zeroth order Hamiltonian, (Eq. \ref{eq:dirac}), with a pre-factor, $\lambda$, referred to as the field strength and proportional to $I_u^M$:
\begin{equation}\label{eq:perturbation}
\hat{H} = \hat{H}_0 + \lambda_u \hat{H}_u^{M,\text{HFS}}.
\end{equation}
An element of the hyperfine coupling matrix can now be calculated as:
\begin{equation}\label{eq:A_uv}
A_{uv}^M = \frac{1}{\langle \tilde{S}_v \rangle} \frac{\text{d} E_{\text{CC}}^{(v)}(\lambda_u)}{\text{d} \lambda_u}.
\end{equation}
The superscript, $(v)$, on the CC energy indicates the quantization axis of the total electronic angular momentum. This axis is in the present work controlled by taking advantage of the symmetry scheme employed by the Dirac program in which (for the symmetries considered here) the quantization axis is fixed along the $z$-axis\cite{Saue1999, DIRAC17release}. $\langle \tilde{S}_v \rangle$ is simply the effective electronic spin and we will denote it $\tilde{S}$.

Due to the axial symmetry in diatomic molecules, the hyperfine interaction tensor can be described in terms of the parallel and the perpendicular components, denoted $A_{\parallel}$ and $A_{\perp}$. If the diatomic molecule is placed along the $z$-axis, $A_{\parallel}$ and $A_{\perp}$ can be calculated as:

\begin{equation}\label{eq:A_par}
A^M_{\parallel} = \frac{1}{ \tilde{S} } \frac{\text{d} E_{\text{CC}}^{(z)}(\lambda_z)}{\text{d} \lambda_z},
\end{equation}
and
\begin{equation}\label{eq:A_perp}
A^M_{\perp} = \frac{1}{ \tilde{S} } \frac{\text{d} E_{\text{CC}}^{(x/y)}(\lambda_{x/y})}{\text{d} \lambda_{x/y}}.
\end{equation}

In practice, the perpendicular component is obtained by placing the internuclear axis on either the $x$- or $y$-axis while the quantization axis of total electronic angular momentum is kept along the $z$-axis, effectively using the expression in Eq. (\ref{eq:A_par}). A similar scheme was recently presented in the framework of the complex generalized Hartree-Fock and Kohn-Sham methods\cite{Gaul2020}. 

\section{Computational details}

All the calculations were carried out with the DIRAC17 program package\cite{DIRAC17release}. In addition to the relativistic 4-component (4c) calculations also the exact 2-component (X2C) method was employed\cite{Ilias2007}. The bond length of the BaF radical was taken from the NIST Chemistry WebBook and has the value of 2.162 \AA \cite{NISTBaF,Knight1971}. For the two isotopes considered in this work, $^{133}$Cs and $^{137}$Ba, nuclear spins of 7/2 and 3/2 and magnetic moments of 2.582$\mu_{\text{B}}$ and 0.937$\mu_{\text{B}}$, respectively, were taken from Ref. \citenum{Stone2005}.

\subsection{Basis sets}

We employ Dyall's relativistic basis sets from the valence, vXz, and core-valence, cvXz, series, where X denotes the cardinal numbers double-, triple-, and quadruple-zeta\cite{Dyall2009,Dyall2012,Dyall2016}. The vXz basis sets include correlation functions (of up to d-, f-, and g-type for Cs and Ba) for the valence region which is defined as 5s5p6s6p. The cvXz basis sets include additional correlation functions (of up to f-, g- and h-type for Cs and Ba) for the core-valence region which includes the 4d shell in addition to the 5s5p6s6p shells. The effect of adding particular types of tight functions, i.e. basis functions with large exponents, was investigated by adding functions in an even-tempered fashion.

\subsection{Correlation treatment}\label{sec:comp_corr}

The unrestricted CC module (RELCC) of DIRAC was employed with different types of perturbative triples\cite{Visscher1996}: the widely used CCSD(T) method\cite{Raghavachari1989} which includes some fifth order triples contributions, the CCSD+T (also called CCSD[T]) method\cite{Urban1985} in which triples contributions only up to the fourth order are included, and the CCSD-T method\cite{Deegan1994} where one further fifth order triples diagram is added to the ones included in the CCSD(T) method\cite{Visscher1996}. The CCSD-T method is therefore formally the most complete method of the three, but its performance was shown to be very similar to CCSD(T)\cite{Deegan1994,Arcisauskaite2012}. In addition we have employed the multi-reference Fock-space CC method (FSCC) \cite{Kaldor1991,Visscher2001}. We have tested the (0,1) sector with varying size of the model space. In sector (0,1) a manifold of singly excited states are obtained by adding an electron to a closed shell singly ionized reference state. The additional electron can occupy those orbitals which are contained in the so-called model space. We will distinguish between two model spaces: A minimum model space (min) only including the valence orbital and an extended model space (ext) which includes the valence orbital as well as the 5 lowest virtual orbitals. 

In both the single-reference CC and the FSCC calculations all electrons were included in the correlation calculation and consequently a high virtual space cut-off of 2000 a.u. was used if not stated otherwise. 

\subsection{Finite field method}

As a consequence of the introduction of the perturbation in Eq. (\ref{eq:perturbation}), the total energy can be written as a Taylor series in $\lambda$:
\begin{equation}\label{eq:Taylor}
E(\lambda) = E^{(0)} + \left. \frac{\partial E(\lambda)}{\partial \lambda} \right| _{\lambda=0} \lambda + \frac{1}{2} \left. \frac{\partial^2 E(\lambda)}{\partial^2 \lambda} \right| _{\lambda=0} \lambda^2 + ... .
\end{equation}

The magnitude of $\lambda$ should be chosen such that higher order terms will be negligible, i.e., $E(\lambda)$ behaves linearly with small variations in $\lambda$. If indeed $E(\lambda)$ is linear with respect to the variations in $\lambda$ the two-point formula can be used to obtain the derivative: 
\begin{equation}\label{eq:num_deriv}
\left. \frac{\partial E(\lambda)}{\partial \lambda} \right| _{\lambda=0} \approx \frac{E(\lambda) - E(-\lambda)}{2\lambda}
\end{equation}

By using this two-point formula any quadratic terms cancel out, resulting in an error proportional to $\lambda^2$, as shown in Ref. \citenum{Pople1968} and Supplementary Information. 
Field strengths should be chosen large enough so that numerical instabilities are avoided and small enough so that higher order terms can safely be neglected. Therefore, a strict convergence criterion of $10^{-12}$ a.u. for the CC amplitudes was used in the calculations. 

\subsection{Procedure}

Since the HFS operator introduced above (Eq. \ref{eq:qm_operator}) is odd with respect to the time-reversal symmetry , it cannot be added directly on the DHF level, which in the DIRAC program is based on the Kramers-restricted formalism (krDHF). Instead, we add the operator on the CC level which uses the unrestricted formalism. Consequently, both spin-polarization as well as correlation effects are accounted for by the CC iterations. In order to disentangle spin polarization and correlation effects we also performed calculations on the Kramers-unrestricted DHF level (kuDHF) using the ReSpect program\cite{ReSpect}. For a description of the kuDHF method we refer to Ref. \citenum{Malkin2011,Gohr2015,Haase2018}.

For clarity we outline the procedure of the calculation below. We note that the finite field scheme has long been available in the DIRAC program but hasn't, to our knowledge, been applied to HFS constants. In order to construct the HFS operator we simply employ operators from the catalogue of one-electron operators included in the DIRAC program. The scheme is as follows:
\begin{enumerate}
\item Perform an unperturbed Kramers-restricted DHF calculation.
\item Carry out the integral transformation including integrals over the HFS operator, Eq. (\ref{eq:qm_operator}).
\item Determine the DHF energy in the presence of the field from the recomputed Fock-matrix. This will correspond to the Kramers-restricted DHF energy.
\item Perform two Kramers-unrestricted CC calculations in the presence of the positive and negative field to get the field dependent CC energies.
\item Calculate the numerical derivative of the CC energy using the 2-point formula, Eq. (\ref{eq:num_deriv}).
\end{enumerate}


\section{Results and discussion}

\subsection{Numerical accuracy}


Before turning to the effects of basis set, electron correlation, and relativity we devote a section to the investigation of the numerical stability of the scheme presented above. In the case of the the finite field method special care must be taken to avoid numerical instabilities. For this purpose the X2C method and the vdz basis set have been used and only the parallel component, $A_{\parallel}$, of the $^{137}$BaF HFS tensor has been considered as the behavior is expected to be the same for the perpendicular component, $A_{\perp}$.

In order to determine the appropriate field strengths to use with the finite field method, we investigated the dependence of the calculated HFS constants on the field strength. The HFS constants of $^{137}$BaF and $^{133}$Cs on the DHF, CCSD and CCSD(T) level are shown in Tab. \ref{tab:field_strengths} for the field strengths $10^{-9}$, $10^{-8}$, $10^{-7}$, $10^{-6}$, $10^{-5}$, $10^{-4}$, $10^{-3}$, $10^{-2}$, and $10^{-1}$ a.u. In all cases, the results for the lower field strengths of $10^{-9}$, $10^{-8}$, and $10^{-7}$ differ slightly from those obtained with the larger field strengths, indicating numerical instability. Whereas calculations with larger fields all yield the same values of the HFS constant (to the digits shown in the table) at the DHF level, the results on the CC level begin to deviate again at field strengths of $\geq 10^{-2}$. Note that the different dependence of the Hartree-Fock and CC results on the field strengths was also observed and discussed in detail in Ref. \citenum{Pernpointner2001}. The results for field strengths between $10^{-6}$ and $10^{-3}$ are stable for all methods, which indicates that the terms in the Taylor expansion (Eq. (\ref{eq:Taylor})) higher than quadratic are negligible (recalling the cancellation of quadratic terms by the 2-point formula). We have checked this by fitting the total energy as a function of $\lambda$ to a third order polynomial and found that the third order terms only become significant for field strengths above $10^{-3}$ a.u. (see Supplementary Information for further details). From the same fit the error due to neglecting the 3rd order terms (by using the 2-point formula) at field strengths of $10^{-6}$ a.u. can be estimated to be on the order of $10^{-10}$ a.u.. We have thus chosen to use the 2-point formula with a field strength of $10^{-6}$ a.u. for all further calculations. 

\begin{table}[t!]
\begin{center}
\begin{tabular}{cccccccc}
\hline
field & \multicolumn{3}{c}{$^{137}$BaF} & &  \multicolumn{3}{c}{$^{133}$Cs}  \\
\hline
 & ~~DHF~~ & CCSD & CCSD(T) &  & ~~DHF~~ & CCSD & CCSD(T) \\
\cline{2-4}\cline{6-8}
$10^{-9}$ & 1650.2 & 2244.9 & 2244.9 &  & 1500.6  & 2114.8 & 2097.3 \\
$10^{-8}$ & 1644.3 & 2244.9 & 2230.0 &  & 1493.6  & 2110.4 & 2099.0 \\
$10^{-7}$ & 1645.3 & 2247.0 & 2233.3 &  & 1493.0  & 2109.6 & 2097.8 \\
$10^{-6}$ & 1645.2 & 2246.7 & 2233.2 &  & 1493.0  & 2109.5 & 2097.6 \\
$10^{-5}$ & 1645.2 & 2246.7 & 2233.2 &  & 1493.0  & 2109.5 & 2097.7 \\
$10^{-4}$ & 1645.2 & 2246.7 & 2233.2 &  & 1493.0  & 2109.5 & 2097.7 \\
$10^{-3}$ & 1645.2 & 2246.7 & 2233.2 &  & 1493.0  & 2109.5 & 2097.7 \\
$10^{-2}$ & 1645.2 & 2246.4 & 2232.9 &  & 1493.0  & 2109.2 & 2097.4 \\
$10^{-1}$ & 1645.2 & 2216.8 & 2203.1 &  & 1493.0  & 2087.3 & 2075.1 \\
\hline
\end{tabular}
\end{center}
\caption{Calculated $A_{\parallel}$ and $A$ constants (MHz) of $^{137}$Ba in BaF and $^{133}$Cs for different field strengths. The calculations were performed using the X2C method and the vdz basis set.}
\label{tab:field_strengths}
\end{table}

It should be emphasized that the analysis described above should be performed for any new system in consideration. As an example take instead the $^{19}$F HFS constant in BaF, which is around 30 times smaller than the $^{137}$Ba and $^{133}$Cs HFS constants. The range of numerical instability is consequently larger (up to $10^{-6}$ a.u.) for the Ba$^{19}$F results and one would need to use larger field strengths (see Supporting Information). 

To test the numerical accuracy further we have performed a series of tests with the results listed in Tab. \ref{tab:tests}. The first test is related to the dependence of the CC HFS constants on the Hartree-Fock orbitals. We tested two different SCF convergence criteria of $5 \cdot 10^{-9}$ and $1 \cdot 10^{-8}$, resulting in a minor change of 0.05 and $<$0.00 MHz for BaF and Cs, respectively. 

Next we tested the effect of two computational approximations that are commonly employed to speed up the SCF calculations. The first is related to the inclusion of Coulomb integrals. The integrals involving only small-component wave functions, $(SS|SS)$, have in all calculations been replaced by a simple Coulombic correction\cite{Visscher1997a} and the effect of including them is here seen to be -0.24 MHz for both systems. This corresponds to less than 0.02\% of the total values and is similar to that observed in previous studies of contact densities\cite{Knecht2011}. Secondly we tested the effect of screening the two-electron integrals used in the Fock matrix, that is, neglecting those estimated to be below a given threshold\cite{SAUE1997}. A threshold of $10^{-12}$ a.u. is used as default in the DIRAC program and we find that turning the screening off (and thus including all two-electron integrals) has a negligible effect of 0.02 MHz for both systems. 

\begin{table}[t]
\begin{center}
\begin{tabular}{cccc}
\hline
test              &   &   ~~  $^{137}$BaF ~~ & ~~ $^{133}$Cs ~~ \\ 
\hline
SCF convergence & 1e-8    & 2233.59 & 2098.10  \\
                & 5e-9    & 2233.54 & 2098.10  \\ \noalign{\smallskip}
$(SS|SS)$       & exclude & 2233.54 & 2098.10  \\
                & include & 2233.30 & 2097.86  \\ \noalign{\smallskip}
Screening       & 1e-12   & 2233.54 & 2098.10  \\
                & 1e-15   & 2233.56 & 2098.10  \\
                & off     & 2233.56 & 2098.12  \\
\hline
\end{tabular}
\end{center}
\caption{Calculated $A_{\parallel}$ and $A$ constants (MHz) of $^{137}$Ba in BaF and $^{133}$Cs for various computational tests (see text for further details). The calculations were performed using the X2C method and the vdz basis set.
}
\label{tab:tests}
\end{table}

Using field strengths of $10^{-6}$ a.u. and employing the approximations described above, we conclude that we can safely include 4 digits in the following discussions.


\subsection{Basis set} \label{sec:basis_set}

Here we investigate the effect of the basis set on the calculated HFS constants. In order to reach highest possible accuracy we need to choose a basis set which is sufficiently converged with respect to additional functions. We consider the convergence sufficient when additional basis functions don't change the HFS constants by more than $\sim$ 0.5\%, since we expect the total uncertainty of a few percent. At the same time the basis set should be small enough to allow for realistic CC calculations with large active spaces. The following basis set studies were carried out at the 4-component CCSD level correlating all electrons and using a virtual cut-off of 2000 a.u, which will be justified in Section \ref{sec:corr}.


In Tab. \ref{tab:basis_set} the HFS constants of $^{137}$Ba in BaF and $^{133}$Cs are shown with increasing quality of the valence and core-valence basis set series, vXz and cvXz (X = d (double), t (triple), q (quadruple)). For both series and both systems a converging behavior is observed upon increasing basis set quality, with the Cs results converging notably faster than the BaF results. 

The addition of one diffuse function for each angular momentum to the vqz basis set, denoted s-aug-vqz, has negligible effect on the calculated HFS constants. This is as expected since the HFS constants describe the interaction of the unpaired electron with the Ba or the Cs nuclei and thus should not be strongly affected by the quality of the description of the region far away from the nuclei. Note that this is not the case for the HFS constants of excited states, where diffuse functions are of great importance.

The difference between the (c)vtz and (c)vqz results (of approx. 1 \%) indicates however that the basis set is not yet saturated with respect to this property. This can be attributed to the slow basis set convergence of the CC methods\cite{Helgaker2000}. In contrast, previous studies using 4-component DFT methods and the same basis sets showed convergence already at triple-zeta level for the HFS constants\cite{Malkin2011,Haase2018}. 

In Tab. \ref{tab:basis_set} we also show the deviation of the calculated HFS constants from the experimental results\cite{Ryzlewicz1982,Arimondo1977}. For both systems the cvXz HFS constants are higher than the vXz ones, corresponding to a smaller deviation from experiment. On the quadruple-zeta level the difference between the vqz and the cvqz values is $\sim$2\%. The cvXz basis sets include large exponent (tight) functions with high angular momenta, which are needed to correlate the 4d shell (in the case of Ba and Cs) which can be considered as the core-valence region. Since we are correlating all the electrons and considering a property that involves interaction between the valence electrons and the nucleus it is to be expected that core-valence correlation functions are needed for obtaining high accuracy results. 

\begin{table}[t]
\begin{center}
\small
\caption{Calculated $A_{\parallel}$, $A_{\perp}$ and $A$ constants (MHz) of $^{137}$Ba in BaF and $^{133}$Cs for increasing basis set quality. The calculations were performed using the 4C CCSD method. Deviation from the experimental values is also shown.}
\label{tab:basis_set}
\begin{tabular}{lcccccccc}
\hline
 & \multicolumn{5}{c}{$^{137}$BaF}	& & \multicolumn{2}{c}{$^{133}$Cs} \\	
\hline
 &	$A_{\parallel}$	& \%(exp$^a$) & &	$A_{\perp}$	& \%(exp$^a$)	& & $A$	& \%(exp$^b$) \\
 \cline{2-3}\cline{5-6}\cline{8-9}
vdz       & 2247 & -5.4 & & 2168 & -5.8 & & 2110 & -8.2 \\
vtz       & 2316 & -2.5 & & 2238 & -2.7 & & 2206 & -4.0 \\
vqz       & 2342 & -1.4 & & 2264 & -1.6 & & 2232 & -2.9 \\
s-aug-vqz & 2342 & -1.4 & & 2265 & -1.6 & & 2232 & -2.9 \\
cvdz      & 2292 & -3.5 & & 2214 & -3.8 & & 2161 & -6.0 \\
cvtz      & 2363 & -0.5 & & 2285 & -0.7 & & 2264 & -1.5 \\
cvqz      & 2383 &  0.3 & & 2305 &  0.2 & & 2283 & -0.7 \\
aeqz      & 2386 &  0.4 & & 2308 &  0.3 & & 2287 & -0.5 \\ \noalign{\smallskip}
exp  & \multicolumn{2}{c}{2376(12)} & & \multicolumn{2}{c}{2301(9)} & & \multicolumn{2}{c}{2298.16} \\ 
\hline
\multicolumn{4}{l}{\footnotesize{$^a$ Ref. \cite{Ryzlewicz1982} $^b$ Ref. \cite{Arimondo1977}} }
\end{tabular}
\end{center}
\end{table}

\begin{table}[t]
\begin{center}
\small
\caption{Calculated $A_{\parallel}$ and $A$ constants (MHz) of $^{137}$Ba in BaF and $^{133}$Cs with different tight functions added to the vqz basis. The calculations were performed using the 4C CCSD method. The effect (in $\%$) with respect to the vqz basis is also shown. 
}
\label{tab:basis_tight}
\begin{tabular}{lccccc}
\hline \noalign{\smallskip}
 & \multicolumn{2}{c}{$^{137}$BaF} & & \multicolumn{2}{c}{$^{133}$Cs} \\	\hline \noalign{\smallskip}
X &	$A_{\parallel}$	& \%($\frac{\text{X}-\text{vqz}}{\text{vqz}}\cdot100$)	& & $A$	& \%($\frac{\text{X}-\text{vqz}}{\text{vqz}}\cdot100$) \\
\cline{2-3}\cline{5-6} \noalign{\smallskip}
vqz   & 2342 & 0.0 & & 2232 & 0.0 \\
~~+s  & 2342 & 0.0 & & 2231 & 0.0 \\
~~+p  & 2342 & 0.0 & & 2232 & 0.0 \\
~~+d  & 2342 & 0.0 & & 2232 & 0.0 \\
~~+f  & 2366 & 1.0 & & 2262 & 1.4 \\
~~+2f & 2376 & 1.4 & & 2274 & 1.9 \\
~~+3f & 2380 & 1.6 & & 2281 & 2.2 \\
~~+4f & 2383 & 1.8 & & 2285 & 2.4 \\
~~+g  & 2343 & 0.0 & & 2232 & 0.0 \\
~~+h  & 2343 & 0.0 & & 2232 & 0.0 \\
\hline
\end{tabular}
\end{center}
\end{table}

In Tab. \ref{tab:basis_tight} we show the effect of adding tight functions of different symmetries individually to the vqz basis set. Since the behavior of the parallel and perpendicular component of the $^{137}$BaF HFS tensor with respect to basis set is very similar we only considered $A_{\parallel}$ in this case. The conclusion is that only the addition of tight f-functions has an influence on the calculated values. The addition of one tight f-function has the largest effect of 1.0\% and 1.4\% for $^{137}$Ba and $^{133}$Cs, respectively. The addition of another three tight f-functions has a smaller additional effect of 0.8\% and 1.0\% and further tight f-functions are not expected to change the results by more than 0.2\%.

As the cvqz basis set differs from the vqz basis set by 3 tight f-, 2 tight g- and 1 tight h-functions we can conclude that the differences between the vqz and cvqz results are governed by the addition of the tight f-functions. To test that the cvqz is indeed converged with respect to the addition of tight functions we used the all-electron quadruple-zeta basis (aeqz) set which includes correlation functions for all shells, resulting in a minor increase in the HFS constant of $\sim$ 0.2\%. If not stated otherwise we have thus chosen to use the cvqz basis sets in our further investigations of other computational parameters. 

It has been shown previously that the addition of tight s-functions to standard correlation consistent basis sets is necessary to accurately calculate the HFS constants\cite{Hedegard2011,Hedegard2012a}. This is not the case here as seen in Tab. \ref{tab:basis_tight}, indicating that the size of the Dyall vqz basis set in the vicinity of the nucleus is already sufficient.

\begin{table}[t!]
\centering
\small
\caption{Calculated $A_{\parallel}$, $A_{\perp}$ and $A$ constants (MHz) of $^{137}$Ba in BaF and $^{133}$Cs at different levels of correlation. The cvqz basis sets were used in the calculations. 
} \label{tab:corr}
\begin{tabular}{lccccccccccc}
\hline
& \multicolumn{6}{c}{$^{137}$BaF} \\
\hline
 & $A_{\parallel}$ & $\Delta$ corr. & \%(exp$^a$) & $A_{\perp}$ & $\Delta$ corr. & \%(exp$^a$) \\
\cline{2-4}\cline{5-7}
krDHF      & 1598 & 0   & -32.8 &  1553 & 0     & -32.5 \\
kuDHF$^c$  & 1905 & 307 & -19.8 &  1817 & 260   & -21.0 \\
CCSD       & 2383 & 785 & 0.28  &  2305 & 752   & 0.19  \\
FSCCSD min & 2399 & 801 & 0.96  &  2323 & 770   & 0.94  \\
FSCCSD ext & 2403 & 806 & 1.16  &  2328 & 775   & 1.16  \\
CCSD+T     & 2425 & 827 & 2.06  &  2350 & 797   & 2.14  \\
CCSD(T)    & 2358 & 760 & -0.77 &  2282 & 729   & -0.85 \\
CCSD-T     & 2365 & 767 & -0.45 &  2288 & 735   & -0.56 \\
\hline
& \multicolumn{6}{c}{$^{133}$Cs} \\
\hline
&  & &  $A$  & $\Delta$ corr. & \%(exp$^b$) \\
\cline{4-6}
krDHF      & & &  1496 & 0   & -34.9 \\
kuDHF$^c$  & & &  1798 & 302 & -21.8 \\
CCSD       & & &  2283 & 787 & -0.65 \\
FSCCSD min & & &  2302 & 806 & 0.18  \\
FSCCSD ext & & &  2302 & 806 & 0.18  \\
CCSD+T     & & &  2330 & 834 & 1.39  \\
CCSD(T)    & & &  2262 & 766 & -1.58 \\
CCSD-T     & & &  2270 & 773 & -1.24 \\
\hline
\multicolumn{7}{l}{\footnotesize{$^a$ Ref. \citenum{Ryzlewicz1982}, $^b$ Ref. \citenum{Arimondo1977}, $^c$ Results obtained with the ReSpect  } } \\
\multicolumn{7}{l}{\footnotesize{program\cite{ReSpect,Malkin2011}} } 
\end{tabular}
\end{table}

\subsection{Correlation effects} \label{sec:corr}

\begin{figure}[t]
\includegraphics[width=0.6\linewidth]{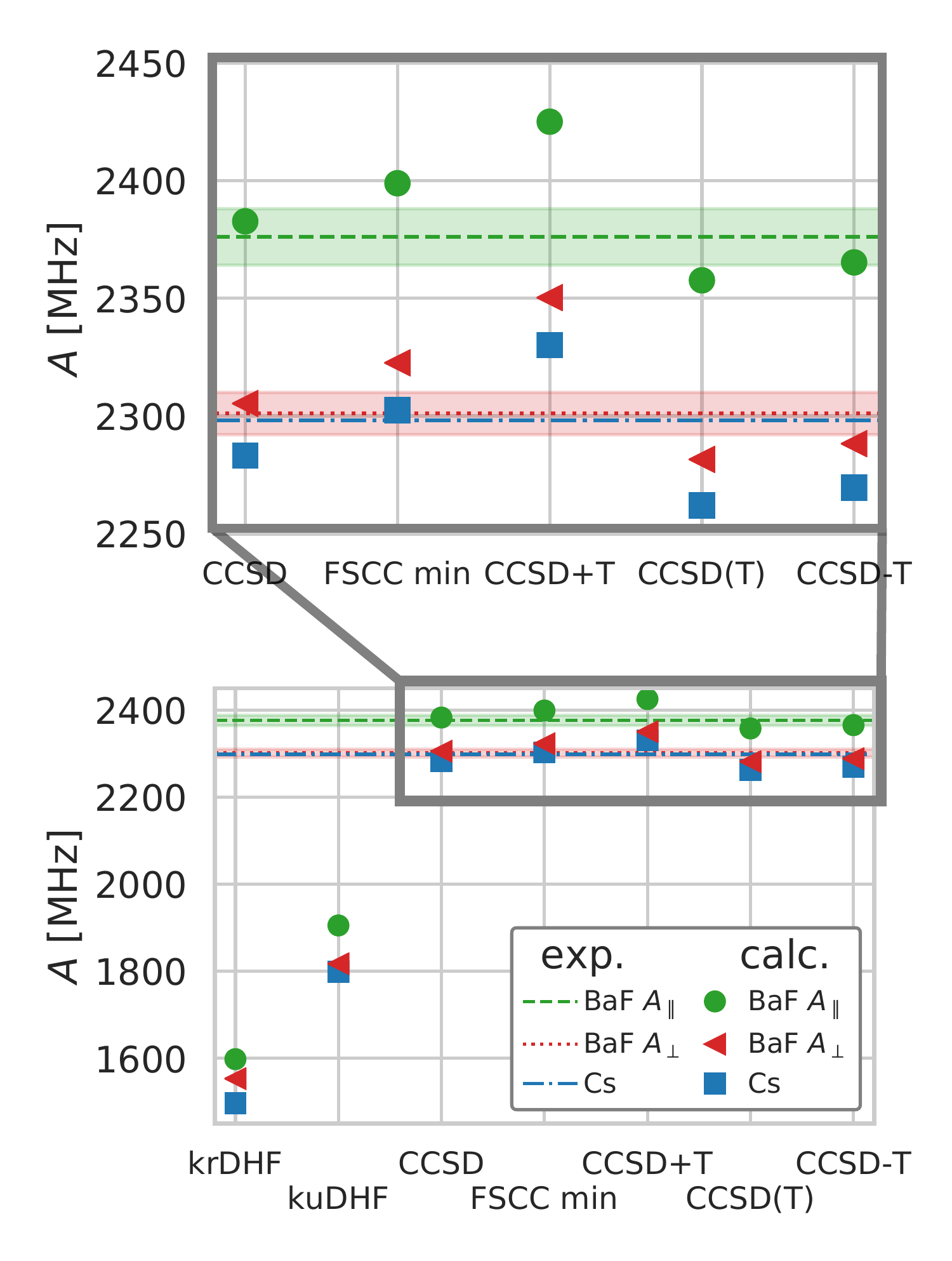}
\caption{Calculated $A_{\parallel}$ and $A$ constants (MHz) of $^{137}$Ba in BaF and $^{133}$Cs at different correlation levels, compared to experiment. The shaded areas indicate experimental uncertainties.
}\label{fig:corr}
\end{figure}

Table \ref{tab:corr} and Figure \ref{fig:corr} contain the HFS constants of $^{137}$BaF and $^{133}$Cs, obtained at different levels of theory. In addition to the total HFS constants, the correlation contribution compared to the krDHF result is shown explicitly along with the deviation from experiment. 

As expected, the lack of correlation treatment as well as of spin polarization in the krDHF method results in a significant underestimation of more than 30\% compared to the experimental results. The inclusion of spin polarization in the kuDHF method leads to a significant increase in the HFS constants resulting in a deviation around 20\%. However, one certainly needs to go to the CC methods for high accuracy. 

With the CCSD method the HFS constants are thus significantly higher, resulting in a deviation from experiment of less than 1\%. The multi-reference Fock-space CC method (FSCCSD) produces results in between the CCSD and CCSD+T values, which is due to the fact that the FSCCSD method takes part of higher order contributions (beyond the double excitations of CCSD) into account due to its  multi-reference formalism. Extending the model space used with FSCCSD (FSCCSD ext, see Section \ref{sec:comp_corr} for a description of the employed model spaces), has a negligible effect, indicating good description of the two systems by a single reference determinant, $^2\Sigma_{1/2}$ in the case of BaF and $^2S_{1/2}$ in the case of Cs. 

The inclusion of perturbative triples has a small effect, with the CCSD+T results slightly overestimating and the CCSD(T) and CCSD-T slightly underestimating the experimental values (see inset of Fig. \ref{fig:corr}).  A similar non-systematic behavior was observed in Ref. \citenum{Arcisauskaite2012} for electric field gradients. However, the present findings are unusual in that the fluctuations in the size of the perturbative triples contributions obtained with the different approximations are comparable with their total values (that is, the difference between the CCSD and CCSD+T/(T)/-T results). For the effective field gradients \cite{Arcisauskaite2012} and in the recent studies of various P- and P,T-odd interaction constants \cite{Hao2018,Denis2019} these fluctuations were significantly smaller than the total contribution of the perturbative triple excitations. 

Our results indicate that the triple excitations are more important for the HFS constants than for the other properties mentioned above. This has been recognized in the past, by, for example, Safronova et al.\cite{Safronova1999}, or more recently by Tang et al.\cite{Tang2019}, who identified this issue from the relatively large difference between the linearized and the full CCSD method. Consequently, we choose to continue our analysis with CCSD and to base our recommended values and uncertainty estimates on this method.  

The correlation contributions to the HFS constants are almost identical for $A_{\parallel}$ in BaF and $A$ in Cs whereas the correlation contribution to $A_{\perp}$ in BaF is slightly lower. It is interesting to note that the trends and differences between the different methods are very similar in BaF and Cs, Fig. \ref{fig:corr}.  This indicates that the two system have a similar electronic structure. In BaF one of the two valence electrons of Ba is participating in the bonding to F leaving a \ce{Ba+} like system, which is iso-electronic to the Cs atom. 

\begin{figure}[t]
\centering
\includegraphics[width=0.6\linewidth]{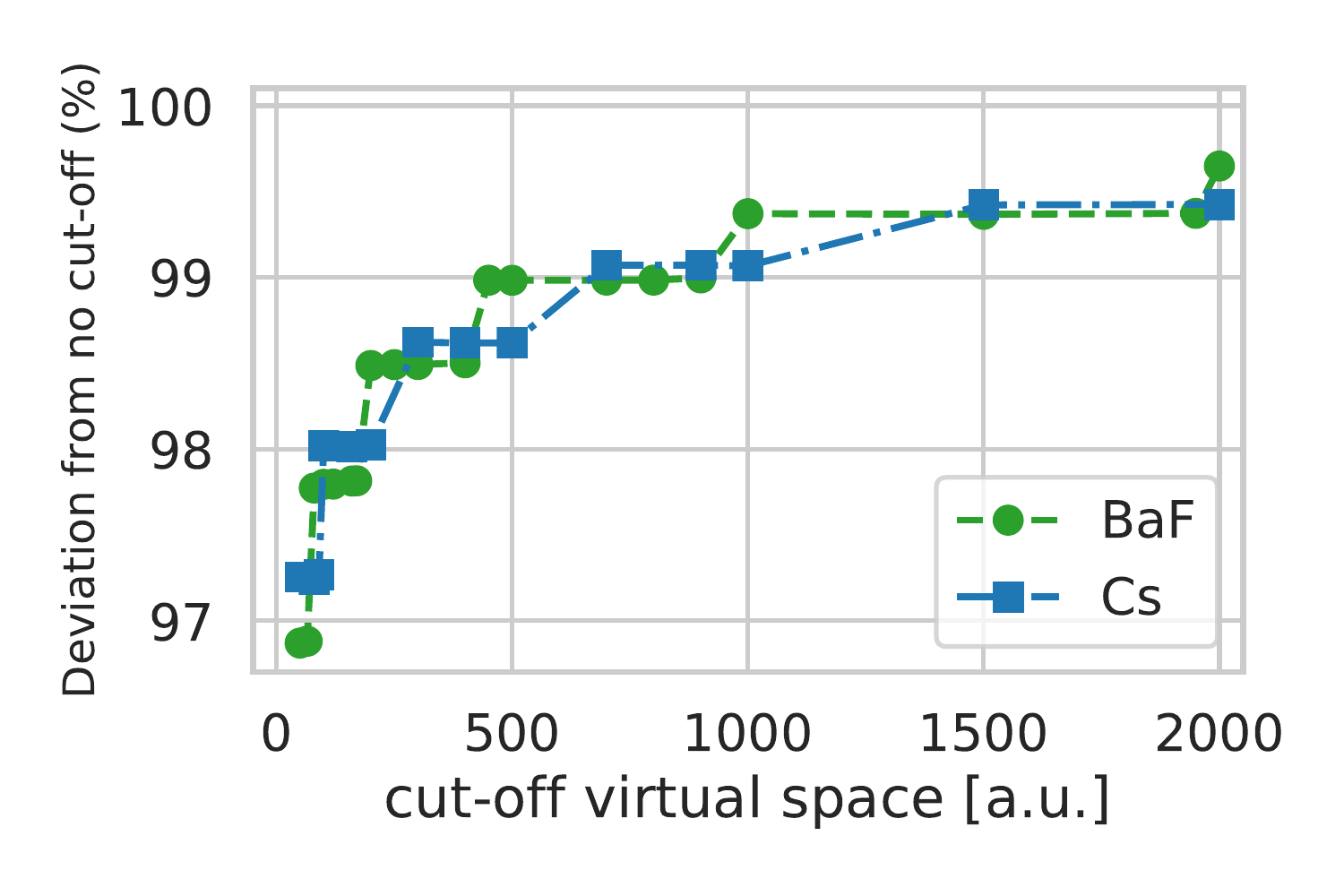}
\caption{Calculated $A_{\parallel}$ and $A$ constants (MHz) of $^{137}$Ba in BaF and $^{133}$Cs at the CCSD/vtz level for different virtual space cut-offs. See text for further details.
} \label{fig:correlation}
\end{figure}

The results presented until now have included correlation of all the electrons and a cut-off of 2000 a.u. of the virtual correlation orbitals. As shown for example in ref. \citenum{Talukdar2018}, a high virtual cut-off is needed in order to capture the correlation contributions to HFS constants associated with the core electrons. In Fig. \ref{fig:correlation} we present in detail the dependence of the HFS constants on the virtual space cut-off when correlating all electrons in BaF and Cs. In both cases only specific virtual orbitals have a significant influence on the correlation contribution to the HFS constants. Inspection of the orbitals in question, see Supporting Information, reveals that the contributing orbitals are all of \textit{s}-function character (s-functions of Ba in the case of BaF). From the deviation with respect to results obtained when all the virtual orbitals were included in the correlation space (designated "no cut-off" on the Fig. \ref{fig:correlation} y-axis) it can be seen that choosing a cut-off of 2000 a.u. will result in an underestimation of the HFS constants of approximately 0.5 \%. Since this uncertainty is smaller than the expected uncertainty of the method we choose to proceed with a cut-off of 2000 a.u.

\subsection{Relativistic effects}

So far we have presented results on the 4-component Dirac-Coulomb (DC) level of theory. The last part of this analysis is dedicated to the investigation of the dependence of the calculated HFS constants on the treatment of relativistic (and related) effects. The results obtained using different models are shown in Tab. \ref{tab:rel}.

As expected, the X2C and DC Hamiltonians give practically identical results, confirming the excellent performance of the former. 

\begin{table}[t!]
\centering
\small
\caption{Calculated $A_{\parallel}$, $A_{\perp}$, and $A$ constants (MHz) of $^{137}$Ba in BaF and $^{133}$Cs at different levels of treatment of relativistic effects. The cv4z basis sets were used in the calculations. } \label{tab:rel}
\begin{tabular}{lccc}
\hline
                    & \multicolumn{2}{c}{$^{137}$BaF} &  $^{133}$Cs   \\
\hline
 & $A_{\parallel}$ & $A_{\perp}$ & \\
CCSD DC & 2383 & 2305 & 2283  \\
CCSD X2C & 2382 & 2305 & 2283  \\
CCSD DCG & 2382 & 2305 & 2282  \\
CCSD PN & 2414 & 2337 & 2312  \\
\hline
\end{tabular}
\end{table}

In the DC Hamiltonian the 2-electron interaction is approximated by the Coulomb potential, which can be considered as a non-relativistic description (it is instantaneous and not Lorentz invariant). For a proper relativistic description of this interaction one needs to turn to the theory of quantum electrodynamics (QED), where one takes into account the finite speed of light resulting in a non-instantaneous interaction. The lowest order one-photon exchange interaction in the static approximation can be derived in the Feynman gauge or the Coulomb gauge, referred to as the Gaunt and Breit interactions, respectively\cite{Dyall2007}. Whereas the Breit interaction is correct to $\mathcal{O}(\alpha^2)$, the Gaunt interaction is correct to $\mathcal{O}(\alpha)$ and simpler to implement and calculate. The current implementation allows us to include the Gaunt interaction on the DHF level (DCG);  these results are shown in Tab. \ref{tab:rel}. We observe a negligible effect of the Gaunt contribution of $\leq$ -1 MHz on the HFS constants. Previous studies on $^{133}$Cs have considered the Gaunt\cite{Kozlov2001} or the full Breit interaction\cite{Safronova1999,Sushkov2001,Derevianko2001,Ginges2017} at different stages of the calculations. For a thorough comparison and discussion of some of these efforts we refer to Ref. \cite{Derevianko2001}. Compared to the majority of the results (4.87 MHz\cite{Derevianko2001}, 5.0 MHz\cite{Kozlov2001} and 6.00 MHz\cite{Ginges2017}) we however predict the wrong sign as well as a too small an effect for the Gaunt interaction contribution, which might be due to several factors: first of all, we calculate the Gaunt contribution on the DHF level only, lacking any Gaunt contribution on the correlated level. Secondly, we employ the restricted DHF formalism, which might lack relaxation effects. Indeed, the negative Breit contribution obtained in Ref. \cite{Safronova1999} was attributed to the neglect of relaxation effects due to the perturbative approach.

Finally, we test the dependence of the HFS constants on the employed nuclear model. In Tab. \ref{tab:rel} we present results obtained using a point-like description of the nuclear charge (PN). Despite the seemingly big physical difference between the point-like and Gaussian description of the nuclear charge, the effect on the calculated HFS constants is relatively small (1.3\% for $A_{\parallel}$ in BaF and $A$ in Cs and 1.4\% for $A_{\perp}$ in BaF). Nonetheless, the Gaussian model should be employed if high accuracy is desired. In previous studies on the DFT level\cite{Malkin2011,Verma2013}, the effect of the finite size of the nuclear charge distribution was found to be $\sim$ 1\% for Zn HFS constants, $\sim$ 1.5\% for Cd HFS constants and as large as $\sim$ 10-15\% for Hg HFS constants.

The authors of Ref. \citenum{Malkin2011} also investigated the effect of a Gaussian description of the nuclear magnetic moment distribution, which turned out to be negligible for lighter elements and as large as $\sim$ 2\% for Hg. This effect was also studied by Ginges et al.\cite{Ginges2017} who found contributions ranging from 0.18(15)\% for $^{133}$Cs to 4.35(131)\% for $^{225}$Ra, which shows that a finite distribution of the magnetic moment should be included if a small uncertainty is desired for the HFS constants of the 6th row elements. The fact that we neglect this effect in the present calculations is one of the main sources of uncertainty, especially for $^{137}$BaF (see Section \ref{sec:uncertainty}).

\subsection{Uncertainty estimation} \label{sec:uncertainty}

\begin{table}[t!]
\centering
\small
\caption{Summary of the sources of uncertainty (MHz) of the calculated $A_{\parallel}$, $A_{\perp}$ and $A$ constants (MHz) of $^{137}$Ba in BaF and $^{133}$Cs.} \label{tab:uncertainty}
\begin{tabular}{lllrrr}
\hline
 \multicolumn{3}{c}{Source} & \multicolumn{2}{c}{$^{137}$BaF}  & $^{133}$Cs \\
\hline
    &  &  &   $\delta A_{\parallel}$ & $\delta A_{\perp}$ & $\delta A$ \\ 
\cline{4-6}
\multicolumn{5}{l}{Basis set} \\
~ & Quality           & cvqz – cvtz      & 20.00  & 20.00  & 19.0   \\ 
  & Tight functions   & aeqz – cvqz      & 3.00   & 3.00   & 4.00   \\  
  & Diffuse functions & s-aug-vqz – vqz  & 0.00   & 1.00   & 0.00   \\ 
\multicolumn{5}{l}{Correlation} \\ 
~ & Higher order      & 2([-T] - [+T])   & -120.00 & -124.00 & -120.00 \\ 
~ & Virtual cut-off   & all – 2000 (vtz) & 8.18   & 8.18$^{c}$   & 12.78  \\ 
\multicolumn{5}{l}{Relativistic effects} \\
~ & Breit             &  & 5.72  $^{a}$ & 5.53  $^{a}$ & 6.00  $^{b}$ \\ 
~ & \qed              &  & -10.01$^{a}$ & -9.68 $^{a}$ & -10.30$^{b}$ \\ 
~ & Bohr-Weisskopf    &  & -39.56$^{a}$ & -38.26$^{a}$ & -7.60 $^{b}$ \\ 
\cline{4-6} \noalign{\smallskip}
\multicolumn{3}{l}{quadratic sum}  & 128.74 & 132.07 & 123.05 \\ 
\multicolumn{3}{l}{\%}             & 5.40   & 5.73   & 5.28   \\ 
\hline
\multicolumn{6}{l}{$^{a}$ Based on $^{135}$Ba$^{+}$ results from Ref. \citenum{Ginges2017}.}\\
\multicolumn{6}{l}{$^{b}$ Taken directly from Ref. \citenum{Ginges2017}. } \\
\multicolumn{6}{l}{$^{c}$ Used $A_{\parallel}$ results. }
\end{tabular}
\end{table}

Based on the investigations presented in the previous sections we consider the results on the CCSD DC / cvqz level to be our recommended values. On this level of theory the convergence with respect to basis set was sufficient and the correlation treatment was the most reliable. 

In addition to the comparison with experimental results we perform an uncertainty analysis based purely on theoretical considerations.  In cases where no experimental data is available a theoretical uncertainty estimate is crucial for direct applications of the calculated properties in experimental research. Here we follow a similar procedure to that in our previous work on symmetry breaking properties\cite{Hao2018,Denis2019}. In this scheme we estimate the error that is introduced by the different approximations employed in the treatment of the basis sets, electron correlation, relativistic effects and nuclear description. These sources of uncertainty are presented in Tab. \ref{tab:uncertainty}, and discussed in the following.

\subsubsection{Basis set}

In section \ref{sec:basis_set} we investigated the effect on the HFS constants of increasing the basis set size in three aspects; the addition of tight functions, diffuse functions and the general quality. We finally chose to use the cvqz basis set and we estimate the uncertainty that is introduced by truncation at the quadruple-zeta level to be not larger than the difference between the cvtz and cvqz results. The effect of adding additional tight (aeqz) and diffuse (s-aug-vqz) functions turned out to be very small but we include them here for the sake of completeness. Adding all three effects together amounts to 23, 24 and 23 MHz for both $A_{\parallel}$ and $A_{\perp}$ in $^{137}$BaF and $A$ in $^{133}$Cs which corresponds to a bit more that 1\%.

\subsubsection{Electron correlation}

In our previous studies we used the spread in the perturbative triples results (i.e. the difference between the CCSD+T and CCSD-T results) times 2 as an estimate for the order of magnitude of the missing higher order correlation contributions\cite{Hao2018,Denis2019}. In both cases this was close to half of the difference between CCSD and CCSD(T). However, in the case of the HFS constants the difference between CCSD+T and CCSD-T is $\sim$ 60 MHZ for both systems, about 3 times larger than the difference between CCSD and CCSD(T). This is an indication that higher order correlation contributions are more important in the case of HFS constant. As a conservative estimate we use again the spread in the perturbative triples results multiplied by 2, which is the major source of uncertainty and contributes $\sim$ 5\% in both cases.

In section \ref{sec:corr} we found that neglecting the virtual orbitals above 2000 a.u. introduces an error of $\sim$0.5\% and we add this contribution to the uncertainty estimate. 

\subsubsection{Relativistic effects (Breit and \qed)}

In order to estimate the magnitude of the higher order relativistic corrections to the 2-electron interaction we rely on previous works and in particular on the recent study by Ginges et al.\cite{Ginges2017} who systematically investigated various contributions to the ground state HFS constants of a few atoms and ions. 

A thorough discussion on the previous calculations of the Breit contribution to the HFS constant in $^{133}$Cs can be found in Ref. \citenum{Derevianko2001} where also the, at the time, most rigorous calculation of the Breit contribution at the level of third order many-body perturbation theory (MBPT) was presented being 4.9 MHz. In the recent study by Ginges et al.\cite{Ginges2017} this contribution was estimated to be 6.0 MHz at the level of the random phase approximation (RPA). We use the larger value of Ginges et al. to estimate the effect of neglecting the Breit interaction.

To our knowledge, no study of the Breit contribution to the $^{137}$BaF HFS constant was published to date. Due to the similar electronic structure and nuclear charge of $^{137}$BaF and $^{133}$Cs the Breit contribution is expected to be similar and we could use the $^{133}$Cs results as an estimate for the effect in $^{137}$BaF. We choose instead to estimate this effect from the result in Ref. \citenum{Ginges2017} for the $^{135}$Ba$^{+}$ HFS constant. The electronic structure of the Ba$^{+}$ ion is a good approximation to that in BaF, where one of the two valence electrons of Ba participates in the bonding to F leaving Ba effectively with a positive charge. The isotope effect on the Breit contribution is negligible. The Breit contribution was determined in Ref \cite{Ginges2017} to be 0.24\% of the total HFS constant of $^{135}$Ba$^{+}$. Taking this to be representative for the $^{137}$BaF HFS constant we estimate the Breit contribution as 5.72 MHz for $A_{\parallel}$ and 5.53 MHz for $A_{\perp}$, which indeed is very similar to that in $^{133}$Cs.

For higher order corrections to the 2-electron interaction one has to turn to quantum electrodynamics (QED) where the lowest order diagrams (beyond Breit) are the single photon one-loop diagrams, namely the vacuum polarization and the self-energy, QED$^{\text{VP+SE}}$.

Two predictions of the \qed contributions to the HFS constant of Cs are available. One is from Sapirstein et al.\cite{Sapirstein2003} of -9.7 MHz, and the other from Ginges et al.\cite{Ginges2017} of -8.8(15) MHz, which agree within the uncertainty provided for the latter. As an estimate we choose the latter value, including the provided uncertainty. For $^{135}$Ba$^{+}$ Ginges et al. predicted -0.38(4)\% which translates to -10.01 and -9.68 MHz for $A_{\parallel}$ and $A_{\perp}$ in $^{137}$BaF.

\subsubsection{Bohr-Weisskopf effect}

Finally we consider the Bohr-Weisskopf effect, which accounts for the finite distribution of the nuclear magnetization compared to a point-like model employed in this work. Again we use the results from Ref. \citenum{Ginges2017} which, unlike the Breit and the \qed effects, turn out to be quite different for the two systems, i.e. -0.18(14)\% for $^{133}$Cs and -1.26(38)\% for $^{135}$Ba$^{+}$. This difference originates from the different nuclear properties of the two isotopes. The similar nuclear properties of the $^{135}$Ba and $^{137}$Ba isotopes results in a similar Bohr-Weisskopf effect\cite{Ginges} and we use the estimate for $^{135}$Ba$^{+}$ in our uncertainty estimate. We note that besides nuclear structure the Bohr-Weisskopf effect also strongly depends on the electronic state of the system, which was recently demonstrated by Prosnyak et al. for Tl\cite{Prosnyak2019}.


\subsection{Comparison with previous studies}

Before we conclude we compare our results with earlier theoretical values and with experimental results. Since the Gaunt contribution was seemingly unreliable, i.e. predicting the wrong sign, and the perturbative triples contributions seemed unreliable due to their relatively large spread, we choose the DC CCSD results (using the cvqz basis set) to be our best estimate for the HFS constant in these two systems, with the associated uncertainties presented in Tab. \ref{tab:uncertainty}.

For both systems the deviation of the DC CCSD results from the experimental values is below 1\% as can be seen in Tab. \ref{tab:BaF} and \ref{tab:Cs}. This deviation is well below the estimated uncertainty of $>$ 5\%, Tab. \ref{tab:uncertainty}. It illustrates the conservative nature of our error estimate, in particular in the higher order correlation corrections, but it is also a result of cancellation between the uncertainties stemming from basis set, correlation and Breit interaction (positive) and the \qed and Bohr-Weisskopf effects (negative). 

\begin{table}[t!]
\centering
\small
\caption{$A_{\parallel}$ and $A_{\perp}$ of $^{137}$Ba in BaF (MHz). } \label{tab:BaF}
\begin{tabular}{lcccc}
\hline
 Method                           & \multicolumn{4}{c}{$^{137}$BaF}  \\
\hline
                                  & $A_{\parallel}$ & \%(exp) & $A_{\perp}$ & \%(exp) \\ \cline{2-5}
 GRECP SCF-EO\cite{Kozlov1997}    & 2264 & -4.71  &	2186 & -5.00  \\
 GRECP RASSCF-EO\cite{Kozlov1997} & 2272 & -4.38  &	2200 & -4.39  \\ \noalign{\smallskip}
 DF RASCI\cite{Nayak2011}         & 2240 & -5.72  &	2144 & -6.82  \\
 DF MBPT\cite{Nayak2011}          & 2314 & -2.61  &	2254 & -2.04  \\ \noalign{\smallskip}
 DC CCSD(this work)               & 2383(129) &  0.29  & 2305(132) & 0.17       \\ \noalign{\smallskip}
Exp \cite{Ryzlewicz1982}          & 2376(12) &        &	2301(9) &        \\
\hline
\end{tabular}
\end{table}



\subsubsection{BaF}

Two previous studies have reported calculations of the $^{137}$BaF HFS constant; these results are presented in  Tab. \ref{tab:BaF}. The first study by Kozlov et al.\cite{Kozlov1997} reported results obtained with the self consistent field (SCF) and restricted active space SCF (RASSCF) methods with and without core-polarization included with the aid of effective operators (EO). The effect of including core polarization ($\sim$ 780 MHz for $A_{\parallel}$ and $\sim$ 740 MHz for $A_{\perp}$) was seen to be very similar to the effect of going from SCF to CCSD discussed in Sec. \ref{sec:corr}. Furthermore, the RASSCF-EO show little difference to SCF-EO which agrees with the small difference between CCSD and FSCCSD. The restricted active space configuration interaction (RASCI) result of Nayak et al.\cite{Nayak2011} is very similar to the (RAS)SCF-EO results, both underestimating the HFS constant by about 5\% compared to the experimental value. The use of MBPT offers a significant improvement compared to the RASCI results.

From the results listed in Tab. \ref{tab:BaF} the present DC CCSD result has the smallest deviation from the experimental value and offers an improvement of accuracy compared to the earlier investigations. 

\begin{table}[t!]
\centering
\small
\caption{$A$ of \ce{Cs} in MHz. All methods employed the 4-component formalism. +B and +G denote the inclusion of the Breit and Gaunt interaction respectively. For the CCSD methods the procedure used to extract the HFS constant is given in parenthesis. 
} \label{tab:Cs}
\begin{tabular}{lll}
\hline
 Method                                                                   & $^{133}$Cs & \%(exp)         \\
\hline  
MBPT$^{a}$+B\cite{Blundell1991}                                                & 2291.00   & -0.31           \\
SDpT+B\cite{Safronova1999}                                               & 2278.5    & -0.85           \\
MBPT$^{a}$\cite{Blundell1991}+B\cite{Derevianko2001}                           & 2295.87   & -0.10           \\
MBPT$^{a}$+OE+G\cite{Kozlov2001}                                               & 2302      & \phantom{-}0.17 \\
CCSDvT\cite{Porsev2010}+B\cite{Derevianko2001}+\qed~\cite{Sapirstein2003}& 2306.6    & \phantom{-}0.36 \\
CCSD (ECC)\cite{Sasmal2015}                                              & 2179.1    & -5.18           \\
CCSD (Z-vector)\cite{Sasmal2017}                                         & 2218.4    & -3.47           \\
MBPT$^{a}$+B+\qed~\cite{Ginges2017}                                            & 2294.4    & -0.16           \\
CCSD (LCCSD)\cite{Tang2019}                                              & 2345.9    & \phantom{-}2.08 \\
CCSD (finite field, this work)                                           & 2283(123) & -0.66           \\
exp\cite{Arimondo1977}                                                   & 2298.16 \\
\hline
\multicolumn{3}{l}{$^{a}$\footnotesize{MBPT has been used as a general term for atomic many-}} \\
 \multicolumn{3}{l}{\footnotesize{body methods. While the MBPT results were all obtained}} \\
 \multicolumn{3}{l}{\footnotesize{using Brueckner orbitals in the evaluation of HFS matrix}} \\
 \multicolumn{3}{l}{\footnotesize{elements (at the RPA level) there are some smaller differen-}} \\
 \multicolumn{3}{l}{\footnotesize{ces between the methods.}}
\end{tabular}
\end{table}

\subsubsection{Cs}
The HFS constant of Cs has been studied extensively due to its relevance for atomic parity violation experiments\cite{Wood1997,Bennett1999}.
Interpretation of such experiments requires sub 1\% accuracy for the theoretical predictions. As can be seen from Tab. \ref{tab:Cs} this goal has been achieved by several groups over the years using various many-body methods\cite{Blundell1991,Safronova1999,Derevianko2001,Kozlov2001,Porsev2010,Ginges2017}. Most of the results with less than 1\% deviation from experiment were obtained with atomic codes, where use of the radial symmetry can practically eliminate basis set errors. Another feature of these results is that they all include a subset of triple excitations as well as estimates for the Breit and/or \qed corrections. Therefore, while the present DC CCSD values have a similar error with respect to experiment, a direct comparison with the earlier high accuracy studies is not meaningful.

In the recent years Sasmal and co-workers have reported the HFS constants of a large set of atoms and molecules on the CCSD level using the extended CC (ECC) and Z-vector frameworks\cite{Sasmal2015,Sasmal2017}. The ECC is uses a variational CC ansatz which allows for calculating HFS constants as expectation values. The Z-vector technique on the other hand is a way to evaluate the energy derivative of non-variational CC energies. Due to the cumbersome truncation scheme in the case of ECC the Z-vector approach is expected to perform better. Indeed, the deviation with respect to experiment is smaller for the Z-vector result compared to the ECC result but still significantly larger than the aforementioned many-body methods. There can be several reasons for this; first of all, these results were obtained with molecular codes which would suffer from similar basis set uncertainties as presented in this work. Secondly the ECC as well as the Z-vector results were obtained with a virtual cut-off of 60 and 40 a.u., respectively. This cut-off corresponds to the first few points in Fig. \ref{fig:correlation}, which indeed leads to an underestimation of $\sim$3\%.
The advantage of the present finite field approach over the ECC and Z-vector methods is that it allows for the inclusion of pertubational triples which in our case provides an important contribution to the uncertainty estimation.

Recently, an additional study on the DC CCSD level was reported by Tang et al.\cite{Tang2019}. In their approach the linearized expression for the CCSD expectation value was employed while the amplitudes were obtained from a CCSD calculation taking all terms into account. The overestimation of $\sim$ 2\% was attributed to the missing non-linear terms in the expectation value expression.


\section{Conclusion}

We calculated the HFS constants of $^{137}$BaF and $^{133}$Cs on the relativistic coupled cluster level using the finite-field method as a straightforward way to evaluate the energy derivative. This scheme has been previously applied to various properties but the present work is the first application to HFS constants. Consequently, a detailed investigation of computational parameters has been performed and presented. The effect of including different types of perturbative triples on the calculated HFS constants was seen to be more irregular than in the previous studies. We thus expect triple excitations to be important and conclude that a perturbational treatment is insufficient. 

Based on the computational investigations, a transparent theoretical uncertainty estimate has been performed. Because of the irregular behavior of the perturbative triples, the largest contribution to the uncertainty estimate comes from the higher order correlations. Higher order relativistic as well as nuclear magnetization distribution effects were included in the estimate by using results from the literature. The estimated uncertainties amounted to 129 MHz (5.4\%) and 132 MHz (5.7\%) for $A_{\parallel}$ and $A_{\perp}$ in $^{137}$BaF and 123 MHz (5.28\%) for $^{133}$Cs. These uncertainties are notably larger than those predicted for the \textit{P,T}-odd interaction constants ($\sim 2\%$) that were obtained using the same scheme as in the present work\cite{Hao2018,Denis2019,Denis2019a}.

The estimated uncertainties were found to be well above the deviation from experimental results which for both systems was below 1\%. This discrepancy is partly due to the conservative nature of the uncertainty estimate (especially in the case of the higher order correlation effects) but it also reflects a fortunate cancellation of the missing contributions. An important task for the future is consequently to improve the description of higher order correlations which would enable more reliable uncertainty estimates. 

\section*{Acknowledgements}
M.I. acknowledges the support of the Slovak Research and Development Agency and the Scientific Grant Agency, APVV-15-0105 and VEGA  1/0562/20, respectively. This research used resources of a High Performance Computing Center of the Matej Bel University in Banska Bystrica using the HPC infrastructure acquired in projects ITMS 26230120002 and 26210120002 (Slovak infrastructure for high performance computing) supported by the Research and Development Operational Programme funded by the ERDF.
P.A.B.H. and A.B. would like to acknowledge the Center for Information Technology of the University of Groningen for their support and for providing access to the Peregrine high performance computing cluster. P.A.B.H. would like to thank L. Visscher, H. J. Aa. Jensen and M. Repisky for useful discussions.

\providecommand{\latin}[1]{#1}
\makeatletter
\providecommand{\doi}
  {\begingroup\let\do\@makeother\dospecials
  \catcode`\{=1 \catcode`\}=2 \doi@aux}
\providecommand{\doi@aux}[1]{\endgroup\texttt{#1}}
\makeatother
\providecommand*\mcitethebibliography{\thebibliography}
\csname @ifundefined\endcsname{endmcitethebibliography}
  {\let\endmcitethebibliography\endthebibliography}{}

\clearpage

\end{document}